\documentclass[prl,twocolumn,superscriptaddress,showpacs]{revtex4}
\usepackage{graphicx}

\newcommand{\da}{\Delta\alpha/\alpha}
\newcommand{\lsim}{\mbox{$\:\stackrel{<}{_{\sim}}\:$} }

\begin{document}
\bibliographystyle{apsrev}

\preprint{}

\title{Further Evidence for Cosmological Evolution of the Fine
Structure Constant}

\author{J.K. Webb}
\affiliation{School of Physics, University of New South Wales, Sydney,
NSW 2052, Australia}
\author{M.T. Murphy}
\affiliation{School of Physics, University of New South Wales, Sydney,
NSW 2052, Australia}
\author{V.V. Flambaum}
\affiliation{School of Physics, University of New South Wales, Sydney,
NSW 2052, Australia}
\author{V.A. Dzuba}
\affiliation{School of Physics, University of New South Wales, Sydney,
NSW 2052, Australia}
\author{J.D. Barrow}
\affiliation{DAMTP, Centre for Mathematical Sciences, Wilberforce
Road, Cambridge University, Cambridge CB3 0WA, England}
\author{C.W. Churchill}
\affiliation{Department of Astronomy \& Astrophysics, Pennsylvania\\
State University, University Park, PA 16802, USA}
\author{J.X. Prochaska}
\affiliation{Carnegie Observatories, 813 Santa Barbara Street,
Pasadena, CA 91101, USA}
\author{A.M. Wolfe}
\affiliation{Department of Physics and Center for Astrophysics and
Space Sciences, University of California, San Diego, C-0424, La Jolla,
CA 920923, USA}

\date{\today} 

\begin{abstract}
\vskip 0.2in We describe the results of a search for time variability
of the fine structure constant $\alpha$ using absorption systems in
the spectra of distant quasars.  Three large optical datasets and two
21cm/mm absorption systems provide four {\it independent} samples,
spanning $\sim 23$\% to 87\% of the age of the universe.  Each sample
yields a smaller $\alpha$ in the past and the optical sample shows a
$4\sigma$ deviation: $\da = -0.72 \pm 0.18 \times 10^{-5}$ over the
redshift range $0.5 < z < 3.5$.  We find no systematic effects which
can explain our results.  The only potentially significant systematic
effects push $\da$ towards {\it positive} values, i.e. our results
would become more significant were we to correct for them.
\end{abstract}

\pacs{98.80.Es, 06.20.Jr, 95.30.Dr, 95.30.Sf}

\maketitle

A common property of unified theories, applied to cosmology, is that
they allow space and time-dependence of the coupling constants
\cite{theory}.  Spectroscopy of gas clouds which intersect the
sightlines to distant quasars provide stringent constraints on
variation of the fine structure constant $\alpha \equiv
\frac{e^2}{\hbar c}$.  Observing quasars at a range of redshifts
provides the substantial advantage of being able to probe $\alpha$
over most of the history of the universe.

\paragraph*{The many-multiplet method.}
Variations in $\alpha$ would cause detectable shifts in the rest
wavelengths of redshifted UV resonance transitions seen in quasar
absorption systems.  For the relativistic fine structure splitting in
alkali-type doublets, the separation between lines is proportional to
$\alpha ^{2}$, so small variations in the relative separation are
proportional to $\alpha$ \cite{bahcall}.  The ``alkali-doublet'' (AD)
method offers the advantage of being simple, but fails to exploit the
available precision since it compares transitions with respect to the
{\it same} ground state.  In recent papers \cite{dzuba1,webb} we
introduced a new technique, the ``many-multiplet'' (MM) method, which
is far more sensitive than the AD method and which offers other
important advantages.  The MM method allows the simultaneous use of
any combination of transitions from many multiplets, comparing
transitions relative to {\it different} ground-states.  Simultaneously
using species with widely differing atomic masses enhances the
sensitivity because the difference between ground-state relativistic
corrections can be large and even of opposite sign.  The AD method
also fails to fully exploit the available data since only a single
doublet is analysed at a time.  Using several different species at the
same time improves the statistics and, importantly, provides an
invaluable means of minimising systematic effects.

The dependence of the observed wavenumber, $\omega_z$, on $\alpha$ is
conveniently expressed as $\omega_z = \omega_0 + q_1 x + q_2 y$ where
$x=[(\frac{\alpha _z}{\alpha _0})^2-1]$, $y=[(\frac{\alpha _z}{\alpha
_0})^4-1]$, $\alpha_0$ is the present day value, and $\alpha_z$ is the
value at the absorption redshift, $z$.  $q_1, q_2$ are coefficients
which quantify the relativistic correction for a particular atomic
mass and electron configuration.  These coefficients have been
calculated in \cite{dzuba1,dzuba2,dzuba3} using accurate many-body
theory methods.  The accuracy of the laboratory wavenumbers,
$\omega_0$, dictates the precision of $\omega_z$ and hence the
constraints on $\da$.  New high precision laboratory measurements of
many species have been carried out using Fourier transform
spectrographs specifically for the purpose of searching for varying
$\alpha$ \cite{pickering}.

The first application of the MM method \cite{webb} used FeII, MgI and
MgII transitions in 30 absorption systems towards 17 quasars and
yielded an order of magnitude gain over previous AD method
constraints.  The results suggest $\alpha$ may have been smaller in
the past: $\Delta \alpha / \alpha = -1.09 \pm 0.36 \times 10^{-5}$ for
$0.5 < z < 1.6$, where $\da = (\alpha_z - \alpha_0)/\alpha_0$.

\paragraph*{The data.}
In the present work, we have re-analysed our initial sample
\cite{churchill, webb}.  Small changes in the definitions of the
spectral fitting regions and in the selection of systems mean we now
have 28 Mg/FeII systems covering redshifts $0.5 < z_{\rm abs} < 1.8$.
The Mg $q$ coefficients are small compared to those for FeII, so Mg
can be thought to act as an ``anchor'' against which shifts in the
FeII lines can be measured.  This large difference between the $q$
coefficients enabled the dramatic sensitivity increase compared to the
AD method.

We include new data \cite{prochaska}, also obtained using the HIRES
echelle spectrograph on the Keck I telescope.  The spectral resolution
is $\sim 7$ km/s for the entire dataset and the signal-to-noise ratio
per pixel is $\sim 30$ for most of the spectra.  This sample is
dominated by 18 damped Lyman-$\alpha$ absorption systems covering
redshifts $1.8 < z_{\rm abs} < 3.5$ towards 13 quasars but also
includes 3 new Mg/FeII absorption systems.  Two further Keck/HIRES
absorption systems are included \cite{outram,lu}.  The redshift range
is on average higher than the data from \cite{churchill, webb}, so
different transitions are used to constrain $\da$. The transitions
used primarily involve multiplets of NiII, CrII and ZnII.  However,
other transitions (MgI, MgII AlII, AlIII, FeII) are also included.  Al
and Si play an analogous ``anchor'' role to Mg in the lower redshift
sample.

There is an important contrast between the previous Mg/FeII
measurements and these new ones: the NiII, CrII and ZnII $q$
coefficients vary not only in magnitude but also in sign.  Some
wavelengths thus shift in {\it opposite} directions for a given change
in $\alpha$.  This, and the greater difference between the $q$
coefficients (compared to Mg/FeII) provides a further sensitivity
gain.  It also dilutes any possible systematic effects, especially any
associated with wavelength calibration of the data (although careful
tests already eliminate this as a source of significant error
\cite{murphy_sys}).  A summary of all $q$ coefficients and all the
wavenumbers used in our analysis, which are related to the same
reference calibration scale, is given in tables 1 of \cite{murphy_mm,
murphy_si}.

A third large new optical dataset is also included in the present
analysis.  This comprises 21 SiIV absorption doublets towards 13
quasar spectra \cite{prochaska}.

HI 21cm absorption lines can be compared with molecular transitions
detected at mm wavelengths to constrain $g_p \alpha^2$ ($g_p$ is the
proton $g$-factor).  We have re-analysed the data from
\cite{drinkwater}, including additional molecular absorption lines.
This provides two new $\da$ estimates at $z = 0.25$ and 0.68 (see
\cite{murphy_rad}).

\paragraph*{Analysis details.}
The analysis methods used in the present work are as described in
\cite{webb} apart from the following improvements.  $\da$ is now
explicitly included as a free parameter in a multi-parameter fit.
Previously we had varied $\da$ externally.  The velocity width
($b$-parameter) of an absorption line is related to the FWHM of the
gaseous atomic velocity distribution by $b = {\rm FWHM}/1.66$, and
$b^2 = \frac{2kT}{M} + b_{\rm turb}^2$ for an ionic species with mass
$M$. The first term describes the thermal component of the line
broadening at kinetic temperature, $T$, and the second describes a
possible turbulent motion.  $T$ and $b_{\rm turb}$ are also now
included as free parameters, and are not degenerate when there are
$\ge 2$ species in a fit.  Note that $\da$ and $z$ are also not
degenerate when there are $\ge 2$ species in a fit.  We have
re-analysed the MgII and FeII data reported in \cite{webb} using the
modified method, and the two sets of results are statistically
indistinguishable.

As in \cite{webb}, to achieve optimal precision from the data, all
physically related parameters ($z$'s and $b$'s) are tied in the
$\chi^2$ minimisation.  A single $z$-parameter is used for different
corresponding species.  Parameter errors were estimated using the
diagonal terms of the inverse of the Hessian matrix (i.e. the
co-variance matrix) at the best fit solution.  Monte Carlo simulations
verified the reliability of the errors derived in this way.

Rigorous consistency checks are imposed before a fit is statistically
acceptable.  The reduced $\chi^2$ for each fit must be $\sim$1.  Each
fit is carried out in 3 different ways, first assuming thermal
broadening (so $b_{\rm turb} = 0$), secondly assuming turbulent
broadening (so $\frac{2kT}{M} = 0$), and thirdly treating $b_{\rm
turb}$ and $T$ as free parameters.  Variations in $\da$ over the 3
fits must not exceed 1$\sigma$.  Only 2 fits out of the optical
dataset failed this test, which provides a simple robustness check on
the derived velocity structure for each absorption complex.  The final
adopted value was that with the smallest reduced $\chi^2$ (which was,
as expected, in all but 3 cases, the third type of fit above).

\paragraph*{Results.}
We now have 72 individual estimates of $\da$ spanning a large redshift
range, providing the most comprehensive constraints so far obtained.
The 7 solid circles (annotated ``many-multiplet'') in Fig. 1. show the
binned results for the re-analysed absorption systems presented in
\cite{webb} and the new points based on the higher redshift Ni/Cr/Zn
data, a total of 49 points \cite{murphy_mm}.

The hollow triangle (annotated ``alkali-doublet''), illustrates the
average result for the 21 SiIV alkali-doublets \cite{murphy_si}.
Table 1 presents a summary of the results for each sample.  The
overall deviation from $\da = 0$ for the whole optical sample is
significant at the $4.1 \sigma$ level.

The new results for the HI 21cm/mm data are: $\da = (-0.10 \pm 0.22)
\times 10^{-5}$ at $z = 0.25$ and $\da = (-0.08 \pm 0.27) \times
10^{-5}$ at $z = 0.68$, assuming, without justification, constant
$g_p$.  The error for each point includes a component of $0.2 \times
10^{-5}$ to allow for possible spatial and velocity segregation of the
HI and mm absorption.  This could be due to slightly different lines
of sight to the background quasar continuum (at such different
wavelengths), or differences along the same line-of sight, or both.  A
recent analysis \cite{carilli} of the same two absorption systems adds
a systematic error of $1.7 \times 10^{-5}$.  Our value is derived
empirically using measurements of the Galactic interstellar medium
(see fig. 2. of \cite{drinkwater}).

\begin{table}[tb]
\renewcommand*{\arraystretch}{1.3}
\squeezetable
\begin{ruledtabular}
\begin{tabular}{lcccc}
\multicolumn{1}{l}{Sample}
& \multicolumn{1}{c}{Method~~}
& \multicolumn{1}{c}{~N$_{\rm abs}$}
& \multicolumn{1}{c}{Redshift}
& \multicolumn{1}{c}{$\da$}
\\
\colrule
FeII/MgII      & MM & 28 & $0.5 < z < 1.8~~$ & $-0.70 \pm 0.23$ \\
NiII/CrII/ZnII & MM & 21 & $1.8 < z < 3.5~~$ & $-0.76 \pm 0.28$ \\
SiIV           & AD & 21 & $2.0 < z < 3.0~~$ & $-0.5  \pm 1.3 $ \\
21cm/mm     & radio &  2 & $0.25, 0.68~~$    & $-0.10 \pm 0.17$ \\
\end{tabular}
\end{ruledtabular}
\caption{Summary of results for 4 independent samples.  Values of
$\da$ are weighted means in units of $10^{-5}$.  MM and AD
indicate ``many-multiplet'' and ``alkali-doublet''. N$_{\rm abs}$ is the
number of absorption systems in each sample.}
\end{table}

\begin{figure}[htb]
\includegraphics[angle=270,width=1\columnwidth]{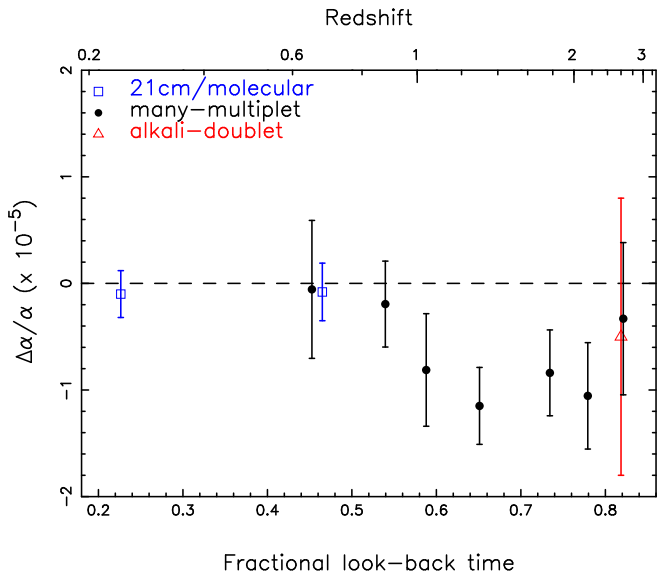}
\vskip 0.1in
\caption{$\da$ vs. fractional look-back time to the Big Bang.  The
conversion between redshift and look-back time assumes H$_0 = 68$
km/s/Mpc, $(\Omega_{\rm M}, \Omega_{\rm \Lambda}) = (0.3, 0.7)$, so
that the age of the universe is 13.9 Gyr.  72 quasar absorption
systems contribute to this binned-data plot. The hollow squares
correspond to two HI 21cm and molecular absorption systems
\cite{murphy_rad}.  Those points assume no change in $g_p$, so should
be interpreted with caution.  The 7 solid circles are binned results
for 49 quasar absorption systems.  The lower redshift points (below $z
\approx 1.6$) are based on (MgII/FeII) and the higher redshift points
on (ZnII, CrII, NiII, AlIII, AlII, SiII) \cite{murphy_mm}. 28 of these
49 systems correspond to the sample used in \cite{webb}.  The hollow
triangle represents the average over 21 quasar SiIV absorption
doublets using the alkali doublet method \cite{murphy_si}.}
\end{figure}

\paragraph*{Potential systematic errors.}
We have carried out a comprehensive search for any systematic effects
\cite{murphy_sys} which could potenatially cause the result we report.
These include: laboratory wavelengths errors, heliocentric velocity
variation during a quasar integration, isotopic saturation and
abundance variation, hyperfine structure, magnetic fields, kinematic
effects, wavelength calibration and air-vacuum wavelength conversion
errors, temperature variations during the observations, line blending,
atmospheric dispersion effects, and variations in the intrinsic
instrumental profile.  None of these are able to explain our result.
For example, kinematic effects (due to velocity segregation for
different species) could introduce a scatter in $\da$ greater than the
statistical error bars, which is not seen.  Only {\it two} potentially
significant systematic effects were identified: atmospheric dispersion
and isotopic abundance evolution.  If the spectrograph slit is not
parallel to the atmospheric dispersion direction (i.e. is not
perpendicular to the horizon), differential dispersion will place the
quasar light at different slit positions, depending on wavelength.  In
fact, this effect turns out to push $\da$ to more {\it positive}
values for each of the 3 optical samples.  If we apply a {\it maximum}
correction, on a case-by-case basis for the actual spectrograph
slit-angle, the result for the MM sample as a whole would become $\da
= (-1.19 \pm 0.17) \times 10^{-5}$.

Quasar absorption system abundances are generally below solar values
\cite{prochaska,churchill}, so isotopic abundance ratios may differ
from terrestrial values.  Therefore, the centroid wavelengths for each
rest-frame transition (from laboratory measurements) may not be quite
correct.  Observations \cite{gay} and theoretical estimates
\cite{timmes} allow us to estimate the importance of this
\cite{murphy_sys}.  To do this we remove all weaker isotopes in all
relevant species and re-fit the entire sample, deriving a new set of
$\da$.  Again, we find that this effect would push $\da$ to more {\it
positive} values for each of the 3 optical samples.  If we were to
apply a correction, we would obtain $\da = (-0.96 \pm 0.17) \times
10^{-5}$ for the whole MM sample.

To summarise the above: ({\it i}) a thorough investigation reveals no
systematic effect which can produce the our results,  ({\it ii})
applying either of the 2 significant corrections would {\it enhance}
the significance of our results.  The results we quote in Table 1 are
not corrected for these systematic effects.

\paragraph*{Other constraints.}

Constraints on $\alpha$ variation come from a variety of independent
sources.  Laboratory measurements made over a 140 day period
\cite{prestage} yield $| \dot \alpha / \alpha | \leq 3.7 \times
10^{-14} ~{\rm yr}^{-1}$.  Another terrestrial constraint comes from
the Oklo natural uranium fission reactor \cite{shylakhter}, active
$\sim 1.8 \times 10^9$ years ago (corresponding to a ``redshift'' of
$z\approx 0.1$).  Recent analyses \cite{damour96, fujii} suggest $\da
= (-0.4 \pm 1.4) \times 10^{-8}$ (although a second, significantly
non-zero solution is also permitted adopting a different Sm resonance
level shift).  The limit above (favoured by \cite{fujii}) is well
below our detection.  The discrepancy is easily removed for a
non-linear time-evolution in $\da$ since the quasar data probe an
earlier epoch.  Note that Fig. 1 shows that our data are consistent
with no variation for $z \lsim 1$.  One may also interpret the
combination of the Oklo and quasar results as the absence of temporal
variation and the existence of spatial variation of $\alpha$.  Also,
unlike the optical quasar data, the Oklo data do not constrain $\da$
directly, but constrain $e^2/r_0 \sim \alpha m_\pi c^2$ ($r_0$ is the
nucleon-nucleon separation, and $m_\pi$ is the $\pi-$meson mass).
Even then, this relies on the unjustified assumption that the strong
interaction and nucleon kinetic energies are constant.  The Oklo
result is thus not as ``clean'' as the quasar results and a reliable
interpretation of the apparent discrepancy requires further work.

Interesting limits can be obtained by comparing the hyperfine 21cm HI
transition with optical atomic transitions in the same gas cloud.
Defining $X = \alpha^2 g_p m_e/m_p$ ($m_e/m_p$ is the ratio of
electron and proton masses), a $z_{abs} = 1.8$ gas cloud provides a
limit of $\Delta X/X = 0.7 \pm 1.1 \times 10^{-5}$ (95\% confidence
limit) \cite{cowie}.  Comparison with our result constrains any
variation of $W = g_p m_e/m_p$ and would give a new result of $\Delta
W/W = 2.1 \pm 0.7 \times 10^{-5}$ (68\% limits).  However, the error
quoted in \cite{cowie} on $\Delta X/X$ does not include any component
associated with spatial and velocity segregation, which is very likely
to be present when comparing transitions at widely different
frequencies, and will be important for a single measurement.  The true
error on $\Delta W/W$ is therefore probably significantly larger than
this.

The cosmic microwave background (CMB) probes $z \sim 1000$, within
$\sim 10^6$ years of the big bang.  Future experiments \cite{cmb1} may
reach $\da \sim 10^{-2} - 10^{-3}$ \cite{cmb2}, although degeneracy
with any electron mass change may reduce this \cite{kujat}.  The
light element abundances constrain the scale lengths of additional
dimensions at the time of primordial nucleosynthesis ($z \sim 10^8 -
10^9$, a few seconds after the big bang).  The $^{4}$He yield is
sensitive to the (uncertain) electromagnetic contribution to the
neutron-proton mass difference \cite{kolb}.  This problem is avoided
for heavier elements and a recent analysis \cite{bergstrom} yields
$|\da | < 2 \times 10^{-2}$.

Interestingly, independent results are now emerging which support the
trend in $\da$ we find.  The most recent CMB data are consistent with
$\alpha$ being smaller in the past by a few percent \cite{cmb3}.
Also, varying speed of light models, \cite{varyc1}, are appealing
because they may explain the supernovae results for a non-zero
cosmological constant and solve other cosmological problems (e.g. the
horizon, flatness, monopole problems) \cite{varyc2}.  These also
require a smaller $\alpha$ in the past.  We anticipate that further
independent quasar data will provide a definitive check on our
results.

\bigskip

\paragraph*{Acknowledgments}
We thank Juliet Pickering and Anne Thorne for crucial experimental
work, and Ulf Griesmann, Sveneric Johansson, Rainer Kling, Ulf
Litz\'{e}n for similar measurements.  JKW and MTM thank the IoA for
hospitality, where some of this work was done.  JDB acknowledges
support from PPARC, the Royal Society and the UNSW Gordon Godfrey
fund.  CWC acknowledges partial support from NASA NAG5-6399 and NSF
AST9617185. AMW acknowledges partial support from NSF grant
AST0071257.  We thank the John Templeton Foundation for support.

\end{document}